\documentclass[fleqn,10pt]{wlscirep}
\usepackage[utf8]{inputenc}
\usepackage[T1]{fontenc}

\begin{document}

\newcommand{\correction}[1]{ {\color{black} #1} }

\title{
Quantum-anomalous-Hall current patterns and interference in thin slabs of chiral topological superconductors} 

\author[1,2,*]{Daniele Di Miceli}
\author[1,3]{Llorenç Serra}
\affil[1]{Institute for Cross-Disciplinary Physics and Complex Systems IFISC (CSIC-UIB), E-07122
Palma, Spain}
\affil[2]{Department of Physics and Materials Science, University of Luxembourg, 1511 Luxembourg, Luxembourg}
\affil[3]{Department of Physics, University of the Balearic Islands, E-07122 Palma, Spain}

\affil[*]{daniele@ifisc.uib-csic.es}

\begin{abstract}
The chiral topological superconductor, which supports propagating nontrivial edge modes while maintaining a gapped bulk, can be realized hybridizing a quantum-anomalous-Hall thin slab with an ordinary $s$-wave superconductor.
We show that by sweeping the voltage bias in a normal-hybrid-normal double junction, the pattern of electric currents in the normal leads spans three main regimes. 
From single-mode edge-current quantization at low bias, to double-mode edge-current oscillations at intermediate voltages and up to diffusive bulk currents at larger voltages.
Observing such patterns by resolving the spatial distribution of the local current in the thin slab could provide additional evidence, besides the global conductance, on the physics of chiral topological superconductors. 
\end{abstract}

\flushbottom
\maketitle

\thispagestyle{empty}

\section*{Introduction}
The chiral topological superconductor (TSC) is a topologically-nontrivial superconducting state which supports 1D chiral Majorana edge modes (CMEM) while maintaining a full pairing gap in the 2D bulk \cite{Topological_classification_symmetries, Topological_Table, Read-Green, TSC_Review, Rev_TI-TSC}.
A recent proposal \cite{Rev_TI-TSC, Chiral_TSC, Chiral_TSC_Half-Integer_Plateau} for the realization of this exotic phase suggests exploiting 
hybridized
quantum-anomalous-Hall (QAH) insulator heterostructures, inducing superconductivity in the QAH chiral edge modes through proximity coupling
with an $s$-wave superconductor.
3D topological insulators (TIs) like $\text{Bi}_2 \text{Se}_3$, $\text{Bi}_2 \text{Te}_3$ and $\text{Sb}_2 \text{Te}_3$ represent promising materials for such a purpose, because they host nontrivial Dirac-cone shaped gapless surface states within a bulk gap $E_g \approx 0.3$ eV \cite{Next_Gen_TI, Bi2Se3_TIs, Observation_TI}, which is larger than the typical energy scale of the room temperature \cite{Review_Bi2Se3_TIs}.
In a thick slab geometry, these topological boundary modes are well-defined for top and bottom surfaces.
However, when placed in an effective 2D thin film, quantum tunnelling couples the Dirac cones localized on opposite surfaces, opening a mass gap in the energy spectrum \cite{Colloquium_QAH, QAHE_in_MTIs}.
Magnetic ordering induced by doping with transition metals (like V or Cr) can give rise to a QAH state with quantized Hall conductance \cite{Observation_QAH_MTI, Realization_QAH_MTI, Enhancing_QAH_Magnetic_Doping}.

\begin{figure}
    \centering
    \includegraphics[width=0.98\linewidth]{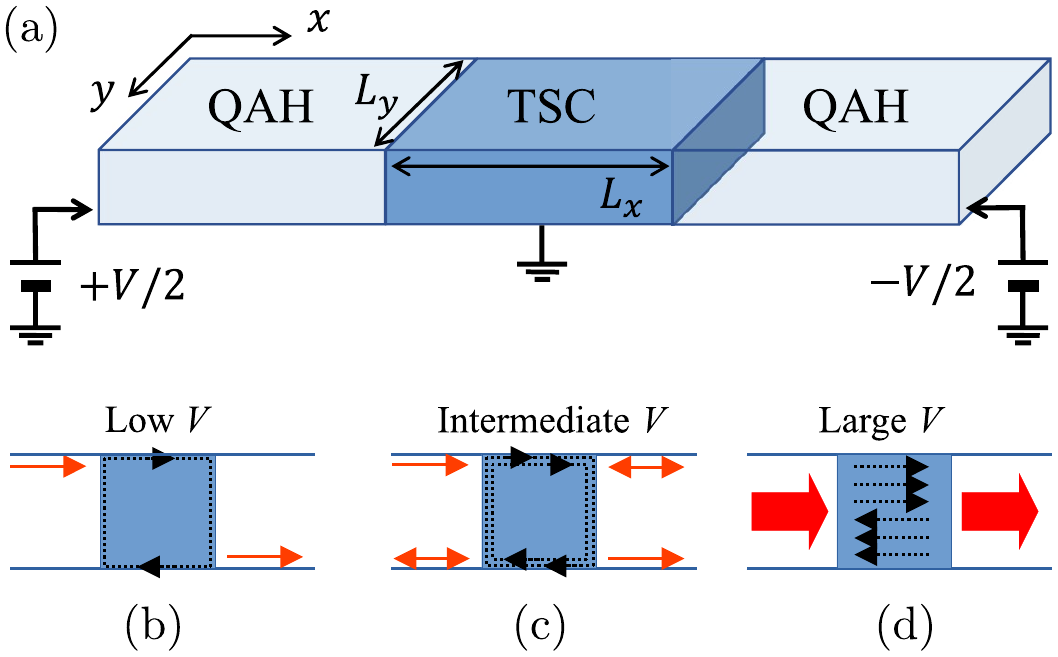} \\
    \caption{
    \label{fig:sketches}
    (a) Sketch of the QAH slab with a TSC hybrid sector proximitized by a superconductor (not shown). 
    (b,c,d) Sketches of the currents (red arrows) for low, intermediate and large applied voltages.
    The dotted black lines represent the TSC quasiparticle modes.}
\end{figure}

In this work, we discuss the electric conductance and the current spatial distribution for the specific case of a QAH thin slab containing a central proximity-hybridised sector in the chiral TSC phase. 
The system is, in practice, a normal-hybrid-normal double junction as sketched in Fig.\ \ref{fig:sketches}a. 
We show here that, when the central sector is a chiral TSC, three different regimes can be distinguished depending on the bias applied across the junction:
    quantized conductance with edge current at low bias (Fig.\ \ref{fig:sketches}b), bias-dependent edge current oscillations at intermediate bias (Fig.\ \ref{fig:sketches}c) and quasiparticle diffusive transport through delocalized states for higher bias exceeding the surface gap (Fig.\ \ref{fig:sketches}d).

\section*{Model}

The 2D minimal model for a full translational invariant TI thin film, which captures the low-energy physics dominated by the Dirac-cone shaped topological surface states (TSSs) around $\mathbf{k}=0$, is given by \cite{Bi2Se3_TIs, Zhang_Hamiltonian, Colloquium_QAH}
\begin{equation}\label{eq:2D_TI_Hamiltonian}
    \mathcal{H}_0 (\mathbf{k}) = \hbar v_F (k_y \sigma_x - k_x \sigma_y) \tau_z + m(\mathbf{k}) \tau_x + \lambda \sigma_z \,,
\end{equation}
in the basis $\psi_\mathbf{k} = (c_{\mathbf{k},\uparrow}^t, c_{\mathbf{k},\downarrow}^t, c_{\mathbf{k},\uparrow}^b, c_{\mathbf{k},\downarrow}^b)$ where $c_{\mathbf{k},\uparrow}^\tau$ annihilates an electron of momentum $\mathbf{k}=(k_x,k_y)$ with spin $\sigma=\uparrow, \downarrow$ on the $\tau=t,b$ layer.
The Pauli matrices $\sigma_{x,y,z}$ $(\tau_{x,y,z})$ act on spin (layer) subspaces, $v_F$ is the Fermi velocity of the Dirac electrons and $m(\mathbf{k})=m_0 + m_1(k_x^2 + k_y^2)$ represents the hybridization between top and bottom TSSs, which at the Dirac point $\mathbf{k}=0$ opens a finite-size gap $m_0$ \cite{Crossover_3D-2D, Massive_Fermions_MTIs}.
Magnetic doping breaks time-reversal symmetry, coupling the electrons to the magnetization through an exchange interaction $\lambda \sigma_z$ \cite{MTIs_Review}.
The topological state of a 2D system with broken time-reversal symmetry is characterized by the Chern invariant \cite{Topological_Table, Topological_classification_symmetries}, which for $|\lambda|>|m_0|$ takes the value $\mathcal{C} = \lambda/|\lambda|$, while for $|\lambda|<|m_0|$ becomes $\mathcal{C}=0$ \cite{QAH_Plateau_Transition}.
The $\mathcal{C}=0$ phase is a trivial-gapped insulator, while the $\mathcal{C}=1$ one corresponds to a QAH state with quantized Hall conductance $\sigma_{xy}=e^2/h$ \cite{TKNN, Colloquium_QAH}.

\subsection*{Superconducting Coupling}
Proximity coupling to an ordinary $s$-wave superconductor induces finite pairing amplitudes $\Delta_{1,2}$ to the electrons in the top and bottom layers of the magnetic TI thin film, respectively \cite{Chiral_TSC, Chiral_TSC_Half-Integer_Plateau}. The effective description of the superconducting system can be carried out through the Bogoliubov de Gennes (BdG) Hamiltonian, which in the basis $\Psi_\mathbf{k} = (\psi_\mathbf{k}, \psi^\dagger_{-\mathbf{k}})$ takes the following form \cite{BdG_Superconductivity, BdG_Methods}
\begin{equation}\label{eq:BdG-Hamiltonian}
    \mathcal{H}_{BdG} (\mathbf{k})  =
    \begin{pmatrix}
        \mathcal{H}_0(\mathbf{k})-\mu & \Delta \\
        \Delta^\dagger & -\mathcal{H}^\star_0(-\mathbf{k})+\mu
    \end{pmatrix} \,, 
\end{equation}
where
\begin{equation}\label{eq:Delta}
    \Delta  = 
    \begin{pmatrix}
        i\Delta_1 \sigma_y & 0 \\
        0 & i\Delta_2 \sigma_y
    \end{pmatrix} 
    \,,
\end{equation}
and $\mu$ represents the chemical potential.
The BdG Hamiltonian preserves  particle-hole symmetry $\mathcal{P}=\delta_x \mathcal{K}$, as $\mathcal{P} \, \mathcal{H}_{BdG}(\mathbf{k}) \, \mathcal{P}^{-1} = -\mathcal{H}_{BdG}(-\mathbf{k})$ by construction.
Here $\delta_{x,y,z}$ stands for a set of Pauli matrices acting in particle-hole space and $\mathcal{K}$ is complex conjugation.
The topological state of the proximitized thin film is thus characterized by an integer topological invariant $\mathcal{N}$ analogous to the Chern invariant \cite{Topological_classification_symmetries, Topological_Table}.
In presence of a finite pairing amplitude, the $\mathcal{C}=1$ QAH state can be viewed as a TSC with $\mathcal{N}=2$ Majorana edge modes \cite{Chiral_TSC, Rev_TI-TSC}, where the doubling of the edge modes is due to the electron-hole degeneracy introduced by the choice of the wavefunction in Nambu space.
Asymmetric pairing amplitudes on top and bottom surfaces \correction{are required to} lift the degeneracy \cite{Chiral_TSC_Half-Integer_Plateau}, leading to a  chiral TSC with $\mathcal{N}=1$ \emph{unpaired} CMEMs propagating over the boundaries.
\correction{For this reason, our numerical results were obtained for fixed $\Delta_1=1.5$ meV and $\Delta_2=0$, meaning that the superconducting pairing acts only on electrons in TSSs on the upper interface of the thin film.
Assuming that the penetration length of Cooper pairs into the TI film is smaller than the thickness \cite{SC_Pairing_TIs}, interlayer pairing terms in Eq.\ \eqref{eq:Delta} can be neglected.}


\subsection*{Edge Picture}

\begin{figure}
    \centering
    \includegraphics[width=\linewidth]{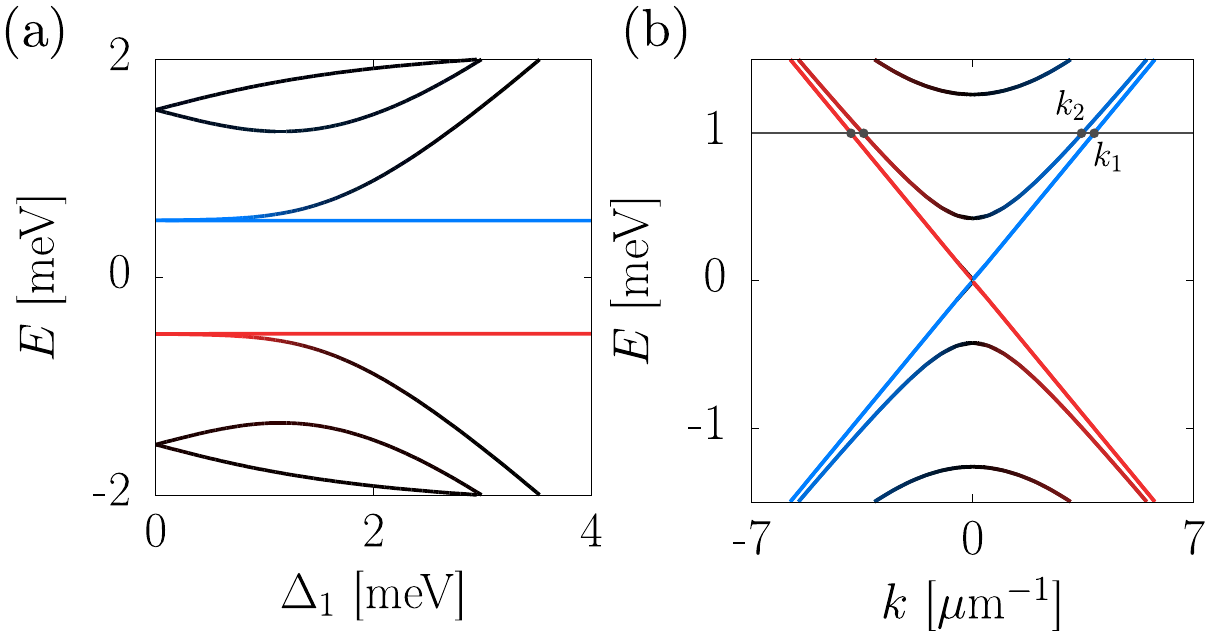}
    \caption{
    \label{fig:modes-localization}
    Low-energy eigenvalues of the BdG Hamiltonian in Eq.\ \eqref{eq:BdG-Hamiltonian} computed for a proximitized magnetic TI thin film with transverse width $L_y=1$ $\mu$m and infinite length.
    In (a) the energy bands are showed as a function of upper pairing amplitude $\Delta_1$ with $k_x=2$ $\mu$m$^{-1}$, while (b) displays the full band structure with $\Delta_1=1.5$ meV. 
    In both plots $\Delta_2=0$, and the other parameters are $\mu=0$, $m_0=1$ meV, $m_1=10^{-3}$ meV $\mu$m$^2$, $\lambda=2$ meV and $v_F=0.26$ meV $\mu$m$/\hbar$.
    Red and blue colours stand for energy modes localized on upper and lower $y$ edges of the slab, respectively.
    \correction{The gray horizontal line is drawn at $E=1$ meV, where a pair of degenerate chiral modes with different wavenumbers $k_1$ and $k_2$ can be found on the same edge of the film.}
    }
\end{figure}

A simple way to understand the emergence of the chiral TSC is to observe the evolution of the edge modes of a QAH thin film under the effect of opposite pairing amplitudes $\Delta_1 = -\Delta_2 = \Delta$ \cite{Rev_TI-TSC, Chiral_TSC}. 
In the QAH phase, the electric current is transported through chiral edge modes, whose $\mathbf{k}=0$ localization length is $\xi \propto 1/|m_0 + \lambda|$ \cite{QSHE_theory-experiment}.
In the simple limit of $\mu = 0$ and in the absence of superconductivity $\Delta=0$, $\mathcal{H}_{BdG}$ reduces to two identical copies of $\mathcal{H}_0$ with opposite sign, thus each chiral edge mode is doubled into a pair of identical electron and hole quasiparticle states  \cite{Colloquium_TIs}.
A finite pairing amplitude couples electron and hole excitations, opening a gap in \emph{one} of the two edge crossings.
In these conditions, the localization length of the doubled edge modes at $\mathbf{k}=0$  can be estimated as $\xi_1 \propto 1/|m_0 + \lambda + \Delta|$ and $\xi_2 \propto 1/|m_0 + \lambda - \Delta|$ \cite{Chiral_TSC, Rev_TI-TSC}, which means that the coupled edge states become delocalized as $\Delta \xrightarrow{} \pm ( m_0 + \lambda )$, moving toward the bulk of the film.
The corresponding eigenstates are pushed to higher energy.

The effect of superconducting coupling is explicitly shown in Fig.\ \ref{fig:modes-localization}a, which displays the low-energy eigenvalues of a magnetic TI thin film for an increasing upper pairing amplitude $\Delta_1$ and $\Delta_2=0$. In the absence of superconductivity, the QAH insulator-superconductor heterostructure hosts a pair of perfectly degenerate edge modes on each side. 
A finite pairing amplitude $\Delta_1$ splits them apart, until a single gapless CMEM remains within the bulk gap, achieving the $\mathcal{N}=1$ phase.
As explained, the second pair of edge crossings becomes gapped and delocalized due to superconducting coupling.
These gapped modes are strongly coupled around $\mathbf{k}=0$, but maintain their edge character away from the Dirac point. 
This fact is shown in Fig.\ \ref{fig:modes-localization}b, which displays the low-energy band structure of an infinite TI thin film with superconducting pairing $\Delta_1=1.5$ meV and $\Delta_2=0$.
As can be noted, at higher energy and for longitudinal wavenumber $k_x \neq 0$, it is still possible to find a pair of chiral modes with \emph{different} wavenumbers $k_1, k_2$ well-localized on the same edge of the system.
\correction{A gray horizontal line is drawn at $E=1$ meV, within the energy range where such a pair of degenerate edge states can be found.}
Due to the spatial interference induced by such a finite momentum difference, the electric current resulting from quasiparticle transport through the edge of a proximitized magnetic TI exhibits characteristic oscillations with respect to propagation length and energy of the incident quasiparticles \cite{Edge-State-Oscillations, Interference_CAES}.

\section*{Electric Transport}

\begin{figure}
    \centering
    \includegraphics[width=0.99\linewidth]{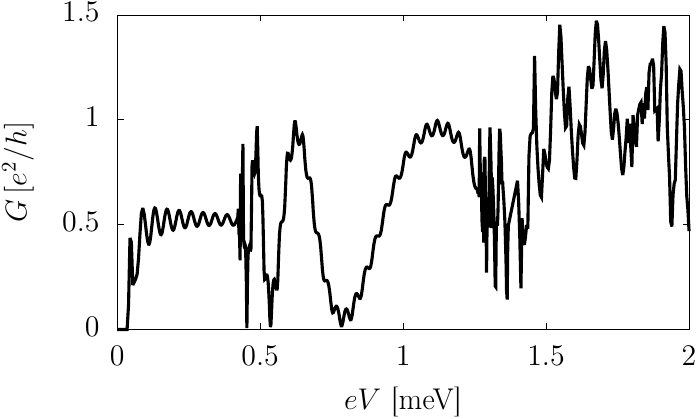}
    \caption{\label{fig:conductance}
    Electric conductance $G$ in the normal leads of a NSN junction as a function of the total bias $V$. 
    We assumed a magnetic TI thin slab of width $L_y=1$ $\mu$m with a central sector long $L_x=20$ $\mu$m.    
    The superconducting pairings are $\Delta_1=1.5$ meV and $\Delta_2=0$, and the other parameters of the effective Hamiltonian are the same as in Fig.\ \ref{fig:modes-localization}.
    }
\end{figure}

We analyse the electric transport through a normal-superconductor-normal (NSN) junction between normal (N) and proximitized (S) magnetic TIs, with the central sector grounded.
The normal leads are held into a QAH phase \cite{Colloquium_QAH}, while the central sector is coupled to a $s$-wave superconductor, such that a pairing amplitude $\Delta_1$ is induced into the TSSs yielding a $\mathcal{N}=1$ phase.

Within the Blonder-Tinkham-Klapwijk formalism, the electric current $I_i$ in the normal terminals $i=1,2$ of a double junction can be expressed as \cite{BTK, Lambert}
\begin{equation}
    I_i = \int_0^{+\infty} dE \sum_a s_a\, \lbrack J_i^a (E) - K_i^a (E) \rbrack \,,
    \label{eq6}
\end{equation}
where we defined the incoming and outgoing fluxes of quasiparticles injected into the superconductor as
\begin{eqnarray}
    J_i^a (E) & = & \frac{e}{h} N_i^a(E)\, f_i^a(E) \,, \\
    K_i^a (E) & = & \frac{e}{h} \sum_{jb} P_{ij}^{ab}(E)\, f_j^b(E) \,.
\end{eqnarray}
Here $a,b \in \{e,h\}$ label electron and hole states, $N_i^a$ is the number of propagating modes in each terminal and $f_i^a$ is the Fermi distribution function.
The fluxes of injected quasiparticles are expressed in terms of the scattering amplitudes $P_{ij}^{ab}$, which represent the transmission probability of a quasiparticle of type $b$ in lead $j$ to a quasiparticle of type $a$ in lead $i$.
Given a voltage drop $V_i$ between the $i$-th terminal and the grounded superconductor, the electric conductance in the normal sectors can be defined as $G_i = I_i / V$, where $V=V_1-V_2$ is the total bias across the junction.
A symmetric bias configuration $V_1 = -V_2$ implies opposite terminal conductances $G \equiv G_1=-G_2$ \cite{Conductance_asymmetry}. 
In terms of scattering amplitudes, the conductance can thus be written as
\begin{equation}
\begin{split}
    G(E) & = \frac{e^2}{2h} \left\lbrack N_1^e(E) - P_{11}^{ee}(E) + P_{11}^{he}(E) \right\rbrack + \\
    & + \frac{e^2}{2h} \left\lbrack P_{12}^{hh}(E) - P_{12}^{eh}(E) \right\rbrack \,,
\end{split}
\end{equation}
where $E=eV/2$ is the energy of the quasiparticles.

The electric conductance $G$ computed in the normal-hybrid-normal double junction is shown in Fig.\ \ref{fig:conductance} as a function of the total bias $V$.
Here, the three different bias-dependent regimes can be clearly distinguished:
\begin{itemize}
\item[a)] a low-bias conductance plateau $G=e^2/2h$, with small oscillations due to finite $L_y$;
\item[b)] an intermediate-bias regime with large conductance oscillations $0<G<e^2/h$;
\item[c)] a metallic-like behaviour at larger biases, with quasiparticle diffusive transport.
\end{itemize}

In the low-bias regime, a single CMEM can be found within the surface gap of the $\mathcal{N}=1$ nontrivial superconductor. 
In a junction with non-proximitized TI films in the QAH state, the electric conductance is expected to be half-quantized at $G=e^2/2h$ \cite{Chiral_TSC_Half-Integer_Plateau, Majorana_backscattering}.
Oscillations around the plateau are due to finite-size coupling between edge states on opposite $y$ sides \cite{CMEM-Oscillations}.
With a higher bias, but still within the surface energy gap, the normal leads remain in the QAH phase, while a pair of chiral edge states with different wavenumbers can be found in the superconducting sector. 
The quasiparticle propagation through the proximitized TI is affected by the spatial interference produced by the phase difference acquired along the propagation length 
$L_x$
\cite{Edge-State-Oscillations, Interference_CAES}, resulting in an oscillatory conductance between $G=0$ (perfect \correction{crossed Andreev reflection}) and $G=e^2/h$ (perfect normal transmission).
Lastly, a further increase in the voltage bias leads to a metallic-like behaviour, due to the activation of delocalized excited states both in the normal and the proximitized sectors.
The quasiparticles transport becomes diffusive rather than ballistic and the electric conductance is expected to be strongly affected by disorder.

\subsection*{Current Distributions}

\begin{figure}
    \centering
    \includegraphics[width=\linewidth]{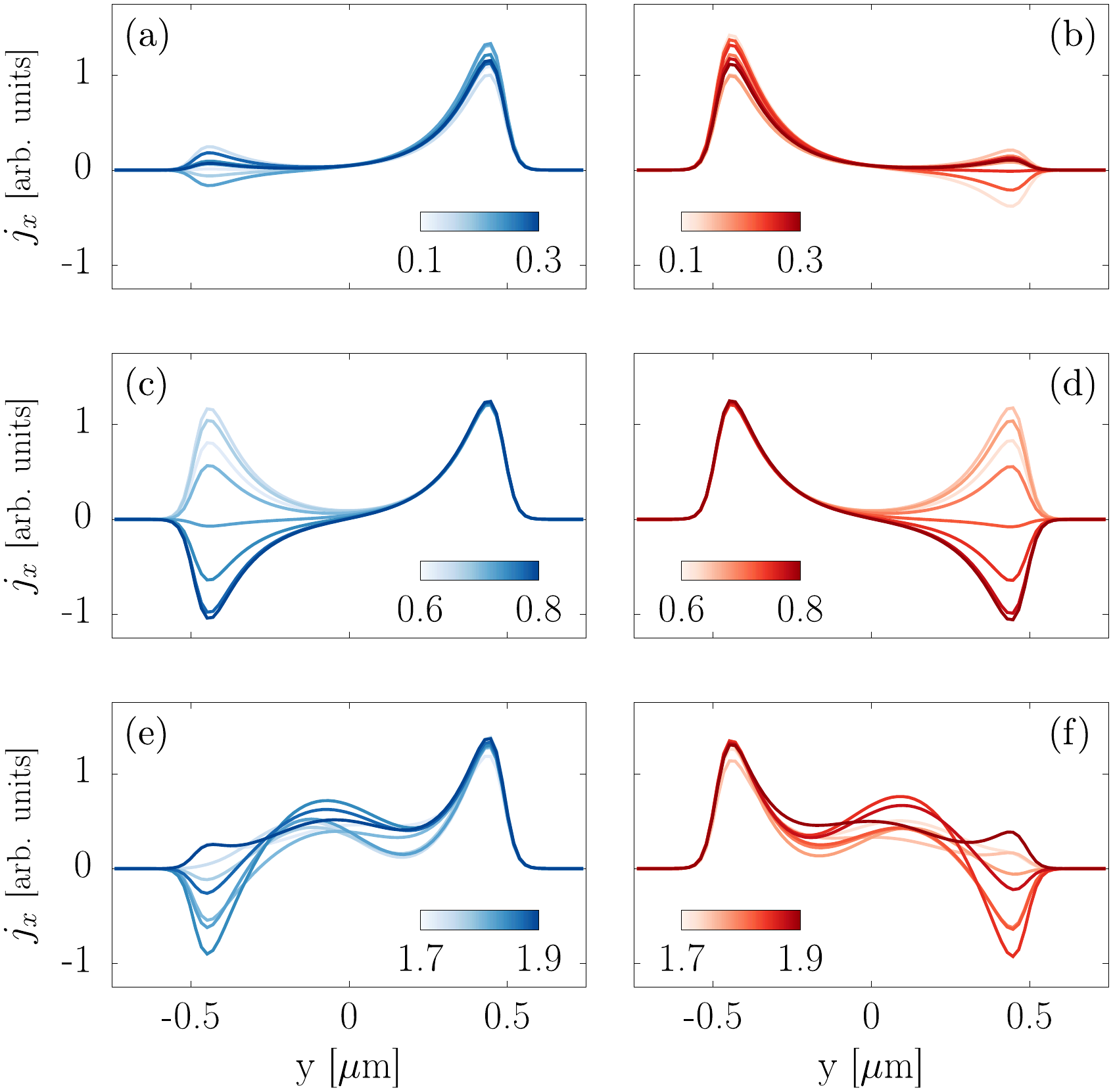}    
    \caption{\label{fig:current-profiles}
    Longitudinal electric-current density $j_x$ computed along the transverse section at $x=-15 \, \mu$m (left column) and $x=15 \, \mu$m (right column) within the two normal leads of the double junction.
    The current density is displayed for a $\mathcal{N}=1$ superconductor in the different regimes: (a)-(b) unpaired CMEMs, (c)-(d) two edge modes regime and (e)-(f) metallic-like state.
    The colour scale indicates the energy $eV$ in meV of the injected quasiparticles.
    Size and parameters of the effective Hamiltonian are the same as in Fig.\ \ref{fig:conductance}.}
\end{figure}

The above different regimes can be further characterized in terms of the current densities along the transverse section of the junction.
Resolving the spatial distribution of currents in proximitized TI thin slabs can provide detailed information on the quasiparticle transport processes occurring in the junction.
Conversely, conductance measurements alone have proven unable to provide an unambiguous signature of chiral TSCs in proximitized QAH film \cite{Retracted_CMEMs, Retraction_Note, Absence_evidence, Plateau_without_Majorana}.
Experimentally, the local probing of currents is in principle possible with magnetic imaging techniques using miniaturized squids \cite{SQUID}.

The longitudinal component of electric-current density is \cite{Osca-complex}
\begin{equation}\label{eq:current-density}
    j_x(x,y) = -e\ \Re[\Psi^*(x,y)\, \hat{v}_x\delta_z\,\Psi(x,y)]\;,
\end{equation}
with the quasi-particle velocity operator given by
\begin{equation}
\hat{v}_x\equiv \frac{\partial \mathcal{H}_{BdG}}{\partial p_x} =
\frac{2}{\hbar^2} m_1 p_x \tau_x\delta_z - v_F\,\sigma_y \tau_z\delta_z\; .
\end{equation}
Eq.\ \eqref{eq:current-density} can be straightforwardly evaluated once the real-space wavefunction $\Psi(x,y)$ is obtained by discretizing the original BdG Hamiltonian $\mathcal{H}_{BdG}$ in a 2D lattice. 
Imposing the continuity of the wavefunction at the interfaces between normal and hybridized sectors yields the full wavefunction in the junction for a fixed energy $eV$.

The current distribution along the transverse section in the two \emph{normal} leads of the NSN junction is displayed in Fig.\ \ref{fig:current-profiles} for the three different bias-dependent regimes.
The current density is computed along $y$ at $x=\pm 15$ $\mu$m, assuming that the proximitized sector has dimensions $L_x=20$ $\mu$m and $L_y=1$ $\mu$m and the reference $x=0$ is set in the middle of it.  
However, as long as the normal leads are held into a QAH state, the current density pattern remains unaffected by the particular choice of $x$ at which the measurement is conducted.

Fig.\ \ref{fig:current-profiles}a-b display the density current profiles for a bias $eV \in \lbrack 0.1,0.3 \rbrack$ meV, which ensures a single CMEM on each side of the superconducting sector. 
As expected for a QAH phase \cite{Colloquium_QAH}, the current is well-localized along the $y$ boundaries: the peaks in the current density profile correspond to the injected quasiparticles, localized on opposite sides and propagating in opposite directions.
Apart from small finite-size ($L_y$) effects, no electric current is transmitted or reflected by the superconductor \cite{Majorana_backscattering}, as evidenced by $j_x = 0$ on the edge opposite to the one hosting the injected current.

A different situation is shown in  Fig.\ \ref{fig:current-profiles}c-d, which display the same current profiles for a higher bias $eV \in \lbrack 0.6,0.8 \rbrack$ meV still lower than the surface gap. 
As in the previous case, the normal sectors are in the QAH phase, with edge-localized current and ballistic transport.
Due to the pair of degenerate edge modes in the superconductor, the quasiparticles propagate across the central sector with oscillating probability of normal transmission and \correction{crossed Andreev reflection}, yielding an oscillatory current density on the edge opposite to the one hosting the injected current.
As can be noted, the current density profile is strongly affected by small variations of the bias, as the transmission amplitudes depend on the momentum difference $\delta_k = k_1-k_2$ between the edge channels, which is set by the energy of the quasiparticles.
These bias-induced edge-current reversals are a conspicuous manifestation of the interference of chiral modes in the TSC. 

Finally, Fig.\ \ref{fig:current-profiles}e-f show the metallic-like behaviour corresponding to a bias $eV \in \lbrack 1.7,1.9 \rbrack$ meV, which is larger than the surface gap. 
Both in the normal leads and in the proximitized sector, the quasiparticles can propagate through delocalized modes, which make the transport diffusive rather than ballistic.
Despite some peaks near the edges of the system, the current is transmitted through the whole section of the junction and disorder is expected to play an important role in the electric transport.

A similar analysis of the transverse current density can also be carried out in the proximitized section of the TI thin film, computing the average value $j_x$ in the central sector of the junction.
Qualitatively similar results are to be expected for the characteristic oscillations of the two-modes regime, \correction{even though} in the hybridized sector the quasiparticle current is screened by the condensate.
We clarify that the characteristic oscillatory regime requires the quasiparticle excited states in the superconductor to be mixtures of electrons and holes.
This requirement is always satisfied in the limit of a small chemical potential $\mu = 0$, meaning that the Fermi energy should coincide with the Dirac point of the TSSs.
Although these fine-tuned conditions constitute a limitation for the validity of our analysis, the band structure of the bismuth-telluride-based TI compounds can be engineered \cite{Band-structure-engineering, Tunable_Dirac-cone_TI}, allowing the manipulation of the Dirac surface states without altering the chemical potential of the bulk crystals.

\section*{Conclusion}
In conclusion, we characterized the electric quasiparticle transport through a NSN junction made by magnetic TI thin slabs, where the normal leads are kept into a QAH state while the proximitized sector realizes a $\mathcal{N}=1$ chiral superconductor. 
Three different electric regimes can be observed by increasing the total bias across the junction.
In the limit of $\mu=0$, we predicted peculiar oscillations in the electric conductance due to the coexistence of a pair of degenerate edge modes in the proximitized sector of the junction.
The emergence of this oscillatory regime is strictly related to the physics of the chiral TSC.

\section*{Data availability}
All data generated or analysed during this study are included in this 
published article.

\bibliography{bibliography}

\section*{Acknowledgements}
This project is financially supported by the QuantERA grant MAGMA by MCIN/AEI/10.13039/501100011033 under project PCI2022-132927. 
L.S.~acknowledges support from Grants  No.~PID2020-117347GB-I00 funded by MCIN/AEI/10.13039/501100011033 and No.~PDR2020-12 funded by GOIB.

\end{document}